\documentstyle[12pt]{article}

\def\beq{\begin{equation}}
\def\eeq{\end{equation}}
\def\bea{\begin{eqnarray}}
\def\eea{\end{eqnarray}}

\def\ba{\begin{array}}
\def\ea{\end{array}}

\def\,{\"{U}}
\def\6{\.{I}}

\begin{document}

\baselineskip 0.9cm
\title{\textbf{Supersymmetric solutions of \emph{PT}-/non-\emph{PT}-symmetric
and non-Hermitian Screened Coulomb potential via Hamiltonian
hierarchy inspired variational method}}
\author{Gholamreza Faridfathi, Ramazan Sever\thanks{%
Corresponding author: sever@metu.edu.tr} \\
\emph{\small \emph{Department of Physics, Middle East Technical
University, 06531 Ankara, Turkey}}}
\date{\today}
\maketitle

\begin{abstract}
The supersymmetric solutions of \emph{PT}-symmetric and
Hermitian/non-Hermitian forms of quantum systems are obtained by
solving the Schr\"{o}dinger equation for the Exponential-Cosine
Screened Coulomb potential. The Hamiltonian hierarchy inspired
variational method is used to obtain the approximate energy
eigenvalues and corresponding wave functions.\newline {\textbf{KEY
WORDS}}{\small : Supersymmetric quantum mechanics, Hamiltonian
hierarchy method, exponential-cosine screened Coulomb
potential}\newline {\textbf{PACS}: 03.65.-w, 12.35.Jh, 21.15.-k}
\end{abstract}

\baselineskip0.9cm\bigskip

\newpage
\
\section{\textbf{Introduction}}

\indent\indent In the past few decades, the supersymmetric
approach has been profitably applied to many non-relativistic
quantum mechanical systems [1-5]. The SUSYQM has provided
satisfactory results concerning different non-relativistic quantum
mechanical systems, such as the exactly solvable and partially
solvable potentials [ 6-9]. The exactly solvable potentials can be
understood in terms of a few basic ideas which include
supersymmetric partner potentials, shape invariance and operator
transformations.

\indent Among the interesting problems of the non-relativistic
quantum mechanics which aim to find exact solutions to the
Schrödinger equation for certain potentials of the physical
interest, the screened coulomb potentials have been studied in a
variety branches of physics such as atomic, nuclear and plasma
physics [10-17]. Various types of the screened Coulomb potentials
like the Yukawa, Debye-H\"{u}ckel and exponential-cosine screened
Coulomb (ECSC) potentials are discussed in non-relativistic quantum
mechanics [18-20]. While the screened Coulomb potential which is in
the vector coupling prescription leads to an exactly solvable for
one dimensional Dirac and Klein-Gordon equations [21-23], the
Schr\"{o}dinger equation for these potentials is not exactly
solvable. Perturbative and approximation methods have been applied
to obtain their energy eigenvalues by using hypervirial/shifted
$1/N$ expansion technique, variational approach, Pad\'{e}
approximates, numerical integration and group theoretical approach
[24-32]. The bound state energies of some potentials like the Morse,
P\"{o}schl-Teller and other exponential type potentials are
evaluated through the SUSYQM method by following the
\emph{PT}-symmetric formalism [30-36]. \emph{PT}-symmetric
Hamiltonians satisfy the parity (\emph{P}) and time reversal
(\emph{T}). The eigenvalue spectra of \emph{PT}-symmetric potentials
may be real or complex [37]. If \emph{PT}-symmetry is not
spontaneously broken, the form of spectra is real. For a class of
non-Hermitian Hamiltonians, the concept of pseudo-hermiticity is
valid [38]. In this work, the energy eigenvalues and corresponding
eigenfunctions of \emph{PT}-/non-\emph{PT}-symmetric and
non-Hermitian types of the Exponential-Cosine Screened Coulomb
potential are obtained by using the Hamiltonian Hierarchy method
within the context of \emph{PT}-symmetric quantum mechanics
(\emph{PT}-SQM).

\indent This paper is organized as follows: In section 2, we give a
brief pedagogical review of the Hamiltonian hierarchy method. In
section 3, we apply this method for the Exponential-cosine screened
Coulomb potential. In sections 4 and 5, the method is applied for
the \emph{PT}-/non-\emph{PT}-symmetric and non-Hermitian cases of
this potential. In section 6, the results are discussed as a
conclusion.
\\

\section{\textbf{Hamiltonian hierarchy method}}

\indent\indent The radial Schr\"{o}dinger equation for some
specific potential energies can only be solved analytically  for
the states with zero angular momentum [36, 37]. However, in
supersymmetric quantum mechanics one can deal with the hierarchy
problem by using effective potentials for non-zero angular
momentum states in order to solve the Schr\"{o}dinger equation
analytically. Hamiltonian hierarchy method suggests a hierarchy
problem in the frame of the SUSYQM in which the adjacent members
are the supersymmetric partners that share the same eigenvalue
spectrum except for the missing ground state.

\indent In this method, the first step is to look for an effective
potential similar to the original specific potential and inspired
by the SUSYQM to propose a superpotential, namely $W_{\left(
l+1\right) }(x)$, as an ansatz, where $\left( l+1\right) $ denotes
the partner number with $l=0,1,2...$. Substituting the proposed
superpotential into the Riccati equation,

\begin{equation}
V_{\left( l+1\right) }(x)-E_{(l+1)}^{0}=W_{\left( l+1\right)
}^{2}(x)-\frac{dW_{(l+1)}(x)}{dx},
\end{equation}

\noindent the $\left( l+1\right) $th member of the Hamiltonian
hierarchy can be obtained. As a result, considering the shape
invariance requirement [14], the bound-state energies can be derived
out through the Eq. (1), and the corresponding eigenfunctions
by means of,

\begin{equation}
\Psi_{(l+1)}(x)=N\exp (-\int^{r}W_{(l+1)}(x^{\prime })dx^{\prime }).
\label{e9}
\end{equation}
\
\section{\textbf{Exponential-cosine screened Coulomb potential}}
\indent\indent The cosine screened Coulomb potential is written as,

\begin{equation}
V(r)=-\frac{q}{r}e^{-\lambda r}\cos (\mu r).  \label{e3}
\end{equation}

\noindent Substituting $\cos (\mu r)=\frac{e^{i\mu r}+e^{-i\mu
r}}{2}$ in the above potential, we get,

\begin{equation}
V(r)=-q\frac{e^{-\lambda r}}{r}(\frac{e^{i\mu r}+e^{-i\mu r}}{2}).
\label{e4}
\end{equation}

\noindent or,
\begin{equation}
V(r)=-\frac{q}{2}\left[\frac{e^{\left( i\mu-\lambda \right)
r}+e^{-\left( i\mu+\lambda \right) r}}{r}\right].  \label{e5}
\end{equation}

\noindent To simplify the calculations and for simplicity, let us
take $ q=2 $. Therefore,

\begin{equation}
V(r)=-\frac{e^{-(\lambda-i\mu)r}}{r}-\frac{e^{-(\lambda+i\mu)
r}}{r}. \label{e6}
\end{equation}
This potential can be considered as two separate parts as,
\begin{eqnarray}
V_{1}(r)&=&-\frac{e^{-(\lambda-i\mu)r}}{r},
\end{eqnarray}
and,
\begin{eqnarray}
V_{2}(r)&=&-\frac{e^{-(\lambda+i\mu) r}}{r}.
\end{eqnarray}

\noindent By defining $\lambda-i\mu=\alpha$ and
$\lambda+i\mu=\beta$, the superpotential proposed as an
\emph{ansatz} for the $ V_{1}(r) $ potential becomes,

\begin{equation}
W_{1\left( l+1\right) }(r)=-\left( l+1\right)\frac{\alpha e^{-\alpha r}%
}{1-e^{-\alpha r}}+\frac{1}{l+1}-\frac{\alpha }{2}. \label{e7}
\end{equation}

\noindent According to the Hamiltonian hierarchy method, the
corresponding eigenfunction for this superpotential will be,

\begin{equation}
\Psi _{01}(r)= \left( 1-e^{-\alpha r}\right)^{l+1}
e^{-(\frac{1}{l+1}-\frac{\alpha}{2})r}. \label{e8}
\end{equation}

\noindent Assuming that the radial trial wave function is given by
(10), we replace $ \alpha $ by the variational parameter $ \mu_{1}
$, and as a result,
\begin{equation}
\Psi _{\mu_{1}}(r)= \left( 1-e^{-\mu_{1} r}\right)^{l+1}
e^{-(\frac{1}{l+1}-\frac{\mu_{1}}{2})r}. \label{e8}
\end{equation}

\noindent The variational energy is given by,

\begin{equation}
E_{\mu_{1}}=\frac{\int_{0}^{\infty }\Psi _{\mu_{1}}(r)\left[ -\frac{1}{2}\frac{%
d^{2}}{dr^{2}}-\frac{e^{-\alpha
r}}{r}+\frac{l(l+1)}{2r^{2}}\right] \Psi _{\mu_{1}
}(r)dr}{\int_{0}^{\infty }\Psi _{\mu_{1}}(r)^{2}dr},  \label{e11}
\end{equation}
\\
\noindent The superpotential proposed as an ansatz for the $
V_{2}(r) $ potential is,

\begin{equation}
W_{2\left( l+1\right) }(r)=-\left( l+1\right)\frac{\beta e^{-\beta r}%
}{1-e^{-\beta r}}+\frac{1}{l+1}-\frac{\beta }{2}, \label{e7}
\end{equation}

\noindent and the corresponding eigenfunction for this
superpotential becomes,

\begin{equation}
\Psi _{02}(r)= \left( 1-e^{-\beta r}\right)^{l+1}
e^{-(\frac{1}{l+1}-\frac{\beta}{2})r}. \label{e8}
\end{equation}
Again, assuming that the radial trial wave function is given by
(14), we can replace $ \beta $ by the variational parameter $
\mu_{2} $, and get,

\begin{equation}
\Psi _{\mu_{2}}(r)= \left( 1-e^{-\mu_{2} r}\right)^{l+1}
e^{-(\frac{1}{l+1}-\frac{\mu_{2}}{2})r}. \label{e8}
\end{equation}
The variational energy is given by,

\begin{equation}
E_{\mu_{2}}=\frac{\int_{0}^{\infty }\Psi _{\mu_{2}}(r)\left[ -\frac{1}{2}\frac{%
d^{2}}{dr^{2}}-\frac{e^{-\beta r}}{r}+\frac{l(l+1)}{2r^{2}}\right]
\Psi _{\mu_{2} }(r)dr}{\int_{0}^{\infty }\Psi
_{\mu_{2}}(r)^{2}dr}, \label{e11}
\end{equation}
\

\noindent Thus, by minimizing the energies $E_{\mu_{1}}$ and
$E_{\mu_{2}}$ with respect to the variational parameter $\mu_{1}$
and $ \mu_{2}$, one obtains the best estimate for the energy of
the exponential-screened Coulomb potential.

\noindent As the exponential-cosine screened Coulomb potential is
not exactly solvable, the superpotentials given by Eqs.(9) and
(13) do not satisfy the Riccati equation, but they do satisfy for
effective potentials instead, $V_{1eff}$ and $V_{2eff}$ as,

\begin{equation}
V_{1eff}\left( r\right)
=\frac{\bar{W}_{1}^{2}-\bar{W\prime}_{1}}{2}+E(\bar{\mu_{1}}).
\label{e14}
\end{equation}
and

\begin{equation}
V_{2eff}\left( r\right)
=\frac{\bar{W}_{2}^{2}-\bar{W\prime}_{2}}{2}+E(\bar{\mu_{2}}),
\label{e15}
\end{equation}
\

\noindent where $\bar{W}_{1}=W_{1}(\alpha=\bar{\mu_{1}})$ and
$\bar{W}_{2}=W_{2}(\beta=\bar{\mu_{2}})$. $\bar{\mu_{1}}$ and $
\bar{\mu_{2}}$ are the parameters that minimize the energy
expectation values (12) and (16). They are given by,

\begin{equation}
V_{1eff}\left( r\right)=-\frac{\alpha e^{-\alpha r}}{1-e^{-\alpha
r}}+\frac{l(l+1)}{2}\frac{\alpha^{2}e^{-2\alpha r}}{(1-e^{-\alpha
r})^{2}}+\frac{1}{2}(\frac{1}{l+1}-\frac{\alpha}{2})^{2}+E(\alpha),
\end{equation}

\noindent and

\begin{equation}
V_{2eff}\left( r\right)=-\frac{\beta e^{-\beta r}}{1-e^{-\beta
r}}+\frac{l(l+1)}{2}\frac{\beta^{2}e^{-2\beta r}}{(1-e^{-\beta
r})^{2}}+\frac{1}{2}(\frac{1}{l+1}-\frac{\beta}{2})^{2}+E(\beta),
\end{equation}
\\
\noindent By substituting the values of $ \alpha $ and $ \beta $
in (19) and (20), one can obtain the bound state energies of the
exponential-cosine screened Coulomb potential as,

\begin{equation}
E=-\frac{q}{2}\left[\frac{1}{(l+1)^{2}}+\frac{\lambda^{2}-\mu^{2}}{4}-\frac{\lambda}{l+1}\right].
\end{equation}
\

\noindent It is interesting to notice that for $ \mu=0 $, the
exponential-cosine screened Coulomb potential reduces to the to the
form called Yukawa potential and that Eq.(21) reduces to,

\begin{equation}
E=-\frac{q}{2}\left[\frac{1}{(l+1)}-\frac{\lambda}{2}\right]^{2}.
\end{equation}

\noindent
\\
\section{\textbf{Non-\emph{PT}-symmetric and non-Hermitian exponential-cosine screened Coulomb case}}

\indent\indent The non-PT and non-Hermitian cosine screened
Coulomb potential can be defined as,

\bigskip

\begin{equation}
V(r)=\frac{iq}{r}e^{-\lambda r}\cos (\mu r).  \label{e15}
\end{equation}

\noindent or simply,

\begin{equation}
V(r)=\frac{iq}{2}(\frac{e^{-\alpha r}}{r}+\frac{e^{-\beta r}}{r}).
\label{e16}
\end{equation}

\noindent In this case the proposed superpotentials can be,
\begin{equation}
W_{1\left( l+1\right) }(r)=-\left( l+1\right)\frac{i\alpha e^{-\alpha r}%
}{1-e^{-\alpha r}}+\frac{1}{l+1}-\frac{\alpha }{2}. \label{e7}
\end{equation}

\noindent and

\begin{equation}
W_{2\left( l+1\right) }(r)=-\left( l+1\right)\frac{i\beta e^{-\beta r}%
}{1-e^{-\beta r}}+\frac{1}{l+1}-\frac{\beta }{2}, \label{e7}
\end{equation}

\noindent Though the superpotentials are complex, following the
same method will yield the same energy eigenvalues as in (21).
\\

\section{\textbf{\emph{PT}-symmetric and non-Hermitian exponential-cosine
screened Coulomb case}}

\indent\indent The PT symmetric and non-Hermitian cosine screened
Coulomb potential can be introduced as,
\begin{equation}
V(r)=-\frac{q}{r}e^{-i\lambda r}\cos (\mu r).  \label{e18}
\end{equation}

\noindent or,
\begin{equation}
V(r)=-\frac{q}{2}\left[\frac{e^{-i\left(\lambda-\mu\right)
r}+e^{-i\left(\lambda+\mu\right) r}}{r}\right].  \label{e19}
\end{equation}

\noindent Taking, $\left( \lambda-\mu\right) =-\alpha _{0}$ and
$\lambda+\mu=\beta _{0}$, we will have,

\begin{equation}
V(r)=-\frac{q}{2}(\frac{e^{-i\alpha _{0}r}}{r}+\frac{e^{-i\beta
_{0}r}}{r}), \label{e20}
\end{equation}

\noindent and as a result the superpotentials can be proposed as,

\begin{equation}
W_{1\left( l+1\right) }(r)=-\left( l+1\right)\frac{\alpha_{0} e^{-i\alpha_{0} r}%
}{1-e^{-i\alpha_{0} r}}+\frac{1}{l+1}-\frac{\alpha_{0} }{2}.
\label{e7}
\end{equation}

\noindent and
\begin{equation}
W_{2\left( l+1\right) }(r)=-\left( l+1\right)\frac{\beta_{0} e^{-i\beta_{0} r}%
}{1-e^{-i\beta_{0} r}}+\frac{1}{l+1}-\frac{\beta_{0} }{2},
\label{e7}
\end{equation}
\\
\noindent In conclusion, by applying the method the same energy
eigenvalues will be obtained as in (21).

\section{\textbf{Conclusions and remarks}}

\indent\indent We have applied the the Hamiltonian hierarchy method
within the framework of the SUSYQM formulation by presenting a
superpotential that yields a trial function to calculate the
approximate bound state energies and corresponding eigenfunctions
for the exponential-cosine screened Coulomb potential. We have also
considered its different symmetric forms in our calculations. As the
energy spectrum of the \emph{PT}-invariant complex-valued
non-Hermitian potentials may be real or complex depending on the
parameters, we have clarified that there are some restrictions on
the potential parameters for the bound states in
\emph{PT}-symmetric, or more generally, in non-Hermitian quantum
mechanics. Furthermore, it is shown that the superpotentials, their
superpartners and the corresponding ground state eigenfunctions
satisfy the \emph{PT}-symmetry condition.

\indent Finally, we can add that our approximate yet accurate
results of complexified exponential-cosine screened Coulomb
potential by the justification of the numerical results presented in
Table 1 motivate an appropriate approach to analyze the exactly and
non-exactly solvable potentials. We believe that this method may
increase the number of applications in the study of different
quantum systems.

\section{Acknowledgements}
This research was partially supported by the Scientific and
Technological Research Council of Turkey.

\newpage
\noindent Table 1 : Energy eigenvalues of ECSC as a function of the
screening parameter $\lambda$ for $1\emph{s}$, $2\emph{p}$,
$3\emph{d}$ and $4\emph{f}$ states in Rydberg units of energy.

\begin{center}
\begin{tabular}{ccccc}\hline\hline
\put(65,0){\bf{~State~1\emph{s}}}\\\hline $

\begin{array}{c}
\bf{Screening} \\
\it{\lambda}
\end{array}
$ & $

\begin{array}{c}
\bf{~~SUSYQM}\\
\it{Our~Work}
\end{array}
$ & $

\begin{array}{c}
\bf{~~Hypervirial}\\
\it{Solution}~[40]
\end{array}
$ & $

\begin{array}{c}
\bf{~~NR-QM}\\
\it{Variational}~[20]
\end{array}
$& $
\begin{array}{c}
\newline
\bf{Exact}\\
\it{Numerical}~[39]
\end{array}$\\\hline
\\

~~\tt{0.020}~~&~~~~~~-0.480290~~&~~~~-0.480310~~&~~~~-0.480300~~&-0.480300 \\[0.2cm]
~~\tt{0.050}~~&~~~~~~-0.451810~~&~~~~-0.451800~~&~~~~-0.451820~~&-0.451800 \\[0.2cm]
~~\tt{0.080}~~&~~~~~~-0.424560~~&~~~~-0.424560~~&~~~~-0.424570~~&--------- \\[0.2cm]
~~\tt{0.100}~~&~~~~~~-0.407070~~&~~~~-0.407050~~&~~~~-0.470600~~&-0.407100 \\[0.2cm]
\end{tabular}
\end{center}
\begin{center}
\begin{tabular}{ccccc}\hline\hline
\put(65,0){\bf{State~2\emph{p}}}\\\hline
\\
~~\tt{0.020}~~~~~~&~~~~-0.211800~~~~&~~~~-0.105890~~~~&~~~~~~~~-0.211900~~~~&~~~~~~-0.211900 \\[0.2cm]
~~\tt{0.050}~~~~~~&~~~~-0.162500~~~~&~~~~-0.080400~~~~&~~~~~~~~-0.161500~~~~&~~~~~~--------- \\[0.2cm]
~~\tt{0.080}~~~~~~&~~~~-0.050500~~~~&~~~~-0.046000~~~~&~~~~~~~~---------~~~~&~~~~~~--------- \\[0.2cm]
~~\tt{0.100}~~~~~~&~~~~-0.092860~~~~&~~~~-0.008000~~~~&~~~~~~~~-0.092890~~~~&~~~~~~-0.093070 \\[0.2cm]
\end{tabular}
\end{center}
\begin{center}
\begin{tabular}{ccccc}\hline\hline
\put(65,0){\bf{State~3\emph{d}}}\\\hline
\\
~~\tt{0.020}~~~~~~&~~~~-0.075020~~~~&~~~~-0.037500~~~~&~~~~~~~~-0.075030~~~~&~~~~~~-0.075030 \\[0.2cm]
~~\tt{0.050}~~~~~~&~~~~-0.033620~~~~&~~~~-0.017340~~~~&~~~~~~~~-0.033740~~~~&~~~~~~-0.033830 \\[0.2cm]
~~\tt{0.080}~~~~~~&~~~~-0.009020~~~~&~~~~-0.008000~~~~&~~~~~~~~---------~~~~&~~~~~~--------- \\[0.2cm]
~~\tt{0.100}~~~~~~&~~~~-0.038889~~~~&~~~~---------~~~~&~~~~~~~~---------~~~~&~~~~~~--------- \\[0.2cm]
\end{tabular}
\end{center}
\begin{center}
\begin{tabular}{ccccc}\hline\hline
\put(65,0){\bf{State~4\emph{f}}}\\\hline
\\
~~\tt{0.020}~~~~~~&~~~~-0.028750~~~~&~~~~-0.014700~~~~&~~~~~~~~-0.028970~~~~&~~~~~~--------- \\[0.2cm]
~~\tt{0.050}~~~~~~&~~~~-0.004100~~~~&~~~~-0.003200~~~~&~~~~~~~~---------~~~~&~~~~~~--------- \\[0.2cm]
~~\tt{0.080}~~~~~~&~~~~-0.184500~~~~&~~~~-0.175000~~~~&~~~~~~~~---------~~~~&~~~~~~--------- \\[0.2cm]
~~\tt{0.100}~~~~~~&~~~~-0.018700~~~~&~~~~---------~~~~&~~~~~~~~---------~~~~&~~~~~~--------- \\[0.2cm]
\end{tabular}
\end{center}


\newpage

\begin{thebibliography}{99}

\bibitem{ref1}  F. Cooper, A. Khare, U. Sukhatme, Phys. Rep. {\bf 251}
 (1995) 267.

\bibitem{ref2}  Richard W. Haymaker, A. R. P. Rau, Am. J. Phys. {\bf 54}
 (1986) 928.

\bibitem{ref3}  E. Drigo Filho, R. Maria Ricota, Phys. Lett. A, {\bf 299}
 (2002) 137.

\bibitem{ref4}  L. Gedenshtein, I. V. Krive, Soviet Phys. Usp. {\bf 28}
 (1985) 645.

\bibitem{ref5}  G. L\'{e}vai, in: H.V. von Gevamb (Ed.), Lecture Notes in
Phys., Vol. \textbf{427}, Springer (1993) 427.

\bibitem{ref6}  E. Drigo Filho, Mod. Phys. Rev. A,  {\bf 9}  (1994) 411.

\bibitem{ref7}  E. Drigo. Filho, R. Maria Ricota, Phys. Atom. Nucl. {\bf
61} (1998) 1836.

\bibitem{ref8}  E. Drigo Filho, R. Maria Ricota, Mod. Phys. Lett. A, {\bf
10} (1995) 1613.

\bibitem{ref9}  E. Drigo Filho, R. Maria Ricota, Mod. Phys. Lett. A, {\bf
15} (2000) 1253.

\bibitem{ref10}
I. B. Golberg, R. H. Pratt, J. Math. Phys. \textbf{28} (1987)
1351.
\bibitem{ref11}
O. V. Gabriel, S. Chaudhuri, R. H. Pratt, Phys. Rev. A,
\textbf{24} (1981) 3088.
\bibitem{ref12}
S. J. R. Crossley, Adv. At. Mol. Phys. \textbf{5} (1969) 237.
\bibitem{ref13}
E. R. Vrscay, H. Hamidian, Phys. Lett. A, \textbf{130} (1988) 141.
\bibitem{ref14}
C. K. Au, Y. Aharonov, Phys. Rev. A, \textbf{20} (1979) 2245.
\bibitem{ref15}
C. K. Au, G. W. Rogers, Phys. Rev. A, \textbf{22} (1980) 1820.
\bibitem{ref16}
G. W. Rogers, Phys. Rev. A, \textbf{30} (1984) 35.
\bibitem{ref17}
A. V. Sargeev, A. I. Sherstyuk, Sov. Nucl. Phys., \textbf{39}
(1984) 731.
\bibitem{ref18}
C. H. Mehta, S. H. Patil, Phys. Rev. A, \textbf{17} (1978) 34.
\bibitem{ref19}
M. L. Du, Phys. Lett. A, \textbf{133} (1988) 109.
\bibitem{ref20}
C. S. Lam, Y. P. Varshni, Phys. Rev. A, \textbf{4} (1971) 1875.
\bibitem{ref21}
M. Znojil, Phys. Lett. A, \textbf{102} (1984) 289.
\bibitem{ref22}
F. D. Adame, A. Rodriguez, Phys. Lett. A, \textbf{198} (1995) 275.
\bibitem{ref23}
N. A. Rao, B. A. Kagali, Phys. Lett. A, \textbf{296} (2002) 192.
\bibitem{ref24}
M. M. Panja, R. Dutt, Y. P. Varshni, Phys. Rev. A, \textbf{42}
(1990) 106.
\bibitem{ref25}
A. Chatterjee, Phys. Rev. A, \textbf{35} (1987) 2722.
\bibitem{ref26}
R. Sever, C. Tezcan, Phys. Rev. A, \textbf{35}  (1987) 2725;
\textit{ibid}, \textbf{36} (1987) 1045.
\bibitem{ref27}
G. M. Harris, Phys. Rev., \textbf{125} (1962) 1131.
\bibitem{ref28}
C. R. Smith, Phys. Rev. A, \textbf{134}  (1964) 1235.
\bibitem{ref29}
G. T. Iafrate, L. B. Mendelsohn, Phys. Rev., \textbf{182} (1969)
244.
\bibitem{ref30}
E. R. Vrscay, Phys. Rev. A, \textbf{33}  (1986) 1433.
\bibitem{ref31}
F. J. Rogers, H. C. Graboske, D. J. Harwood, Phys. Rev. A,
\textbf{1}  (1970) 1577 .
\bibitem{ref32}
J. P. Gazeau, A. Maquet, Phys. Rev. A, \textbf{20}  (1979) 727.
\bibitem{ref33}
E. D. Fliho, R. M. Ricotta, Phys. Lett. A, \textbf{269} (2000)
269.
\bibitem{ref34}
B. Bagchi, C. Quesne, Phys. Lett. A, \textbf{300}  (2002) 18.
\bibitem{ref35}
B. Bagchi, F. Cannata, C. Quesne, Phys. Lett. A, \textbf{269}
(2000) 79.
\bibitem{ref36}
G. Faridfathi, R. Sever, Metin Akta\c{s}, \emph{Journal of
Mathematical Chemistry Vol. \textbf{38}, No.4,} (2005) 533-540 .
\bibitem{ref37}
C. M. Bender, S. Boettcher, Phys. Rev. Lett., \textbf{80} (1998)
5243.
\bibitem{ref38}
A. Mostafazadeh, arXiv: math-ph/0110016; \textit{ibid}
math-ph/0203005; \textit{ibid} math-ph/0107001.
\bibitem{ref39}
R. L. Greene, C. Aldrich, Phys. Rev. A, \textbf{14} (1976) 2363.
\bibitem{ref40}
R. Sever, C. Tezcan, Tr. J. of Physics, \textbf{17} (1993)
459-464.

\end{thebibliography}
\end{document}